\documentclass{iucrjournals}

\usepackage{siunitx}
\usepackage{amssymb}
\usepackage{amsmath}
\usepackage{hyperref}
\usepackage{subcaption}
\DeclareSIUnit\angstrom{\text {Å}}

\title{Initial Characterisation of a Prototype TMR Assembly for an Electron-Driven CANS at CERN's CLEAR Facility}

\author[a]{Laurence Wroe \IUCrCemaillink{laurencewroe@gmail.com}\IUCrOrcidlink{0000-0002-6751-1353}}%
\author[b]{Giorgi Kharashvili\IUCrEmaillink{giorgi.kharashvili@psi.ch}}%
\author[c]{Jonas Okkels Birk\IUCrEmaillink{jobo@teknologisk.dk}}%
\author[b]{Federico Vanti\IUCrEmaillink{federico.vanti@daes.pro}}%
\author[a]{Wilfrid Farabolini\IUCrEmaillink{wilfrid.farabolini@cern.ch}}%
\author[b]{Fares Elattab\IUCrEmaillink{resfaelattab@gmail.com}}%
\author[a]{Davide Gamba\IUCrEmaillink{davide.gamba@cern.ch}\IUCrOrcidlink{0000-0002-1985-1847}}%
\author[a]{Torsten Koettig\IUCrEmaillink{Torsten.Koettig@cern.ch}}%
\author[a]{Roberto Corsini\IUCrEmaillink{roberto.corsini@cern.ch}\IUCrOrcidlink{0000-0002-0934-8199}}%
\author[a]{Steinar Stapnes\IUCrEmaillink{steinar.stapnes@cern.ch}\IUCrOrcidlink{0000-0002-0254-8198}}%
\author[b]{Francois Plewinski\IUCrEmaillink{francois.plewinski@daes.pro}}%

\affil[a]{CERN, CH-1211 Geneva-23, Switzerland}
\affil[b]{DAES, CH-1213 Geneva-23, Switzerland}
\affil[c]{Danish Technological Institute, Taastrup, DK - 2630, Denmark}

\begin{document} 
\maketitle 

\begin{synopsis}
This article presents the design, installation, and first experimental testing of a novel, prototype target–moderator–reflector (TMR) assembly for a compact accelerator-driven neutron source (CANS) driven by a \qty{35}{MeV} electron beam.
\end{synopsis}

\begin{abstract}
The Versatile ULtra-Compact Accelerator-based Neutron source (VULCAN) project is developing a compact accelerator-driven neutron source (CANS) optimised for neutron diffractometry in industrial and university settings. Central to VULCAN is a novel target-moderator-reflector (TMR) assembly optimised to convert a \qty{35}{MeV} pulsed electron beam into short neutron pulses (FWHM $\leq$~\qty{20}{\micro s}) in the \qtyrange{1.5}{3.5}{\angstrom} wavelength range. To validate the simulation-driven design process, a prototype TMR was developed for testing at CERN’s CLEAR facility, and this paper presents the design, installation, and results of the first experimental campaign. While moderated neutron pulses were successfully detected, significant discrepancies were observed between the experimental and simulated energy spectra. Potential causes are discussed and recommendations for follow-up measurements are provided.
\end{abstract}

\keywords{Neutron scattering; neutron time of flight; CANS; electron linac}

\section{Introduction}

Neutron scattering provides a powerful, non-destructive probe of matter at the atomic scale, offering deep penetration and unique sensitivity to light elements and isotopes~\cite{Langel2023_NeutronScattering}. It underpins research and innovation across a diverse range of fields from energy and climate to healthcare and advanced manufacturing, ``provid[ing] the basis for innovation, new and better products and, in result, societal well-being''~\cite{LENSReport}.

Most neutron scattering instrumentation are hosted at large scale research infrastructures (LSRIs) that use nuclear research reactors or spallation sources to generate neutron net yields of $\gtrsim$~\qty{1e16}{n/s} and enable the operation of more than 30 instruments in parallel~\cite{ISIS_Practical}. While there are just over 100 LSRIs in operation worldwide~\cite{IAEA_Spallation,IAEA_Reactor}, the demand for neutrons exceeds supply and many instruments are oversubscribed by more than a factor of two. This imbalance is expected to only worsen as ageing reactors are retired without replacement~\cite{ESFRINeutronScatteringFacilities}.

Compact accelerator-driven neutron sources (CANS) are a promising route to expand access to neutron techniques beyond LSRIs. They deliver lower energy ($\lesssim~\qty{100}{MeV}$), lower power ($\lesssim~\qty{10}{kW}$) particle beams to target–moderator–reflector (TMR) assemblies optimised to create tailored neutron spectra for specific instrumentation and applications. While providing significantly lower fluxes than their LSRI counterparts, CANS enable scalability through smaller footprint and capital cost, reduced operational complexity and licensing requirements, lower radioactive waste production, and flexibility to tailor instrumentation and sample environments to specific user needs~\cite{LENSReport}.  

\begin{figure}[ht] %
    \begin{center}
    \includegraphics[width=0.75\textwidth]{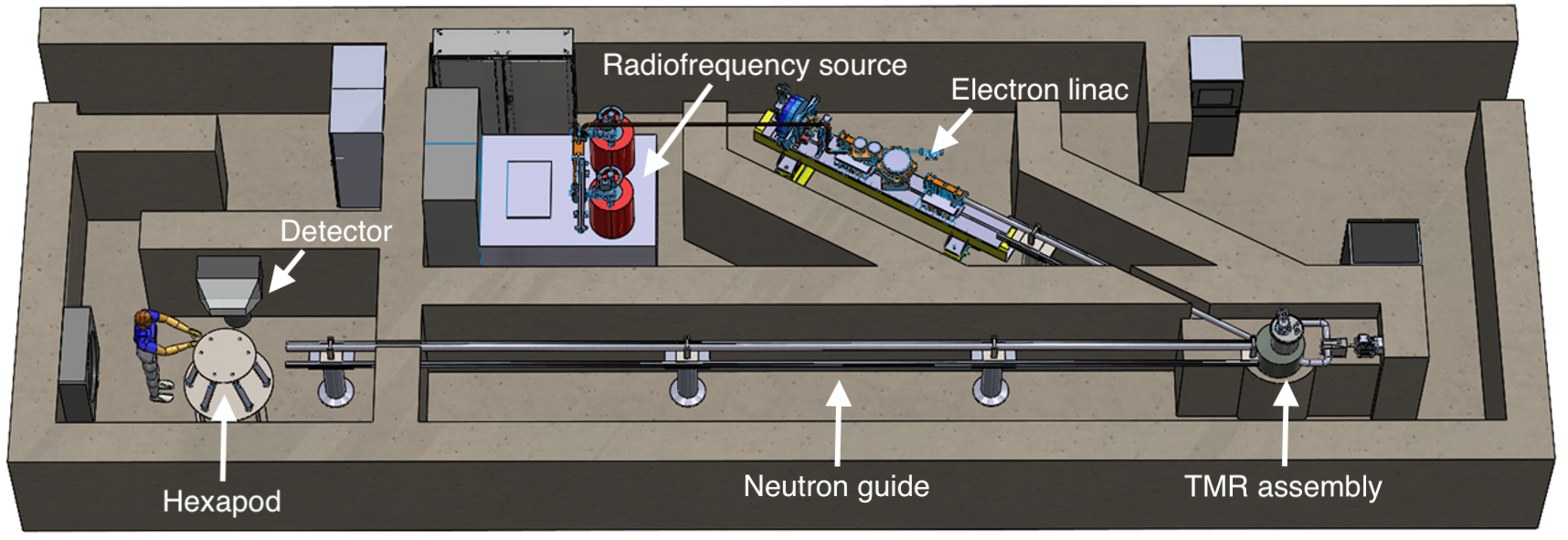} 
    \end{center}
    \caption{Conceptual visualisation of the VULCAN facility.} \label{fig:VULCAN_Schematic}
\end{figure}

The Versatile ULtra-Compact Accelerator-based Neutron source (VULCAN) project seeks to realise a turnkey CANS that is optimised for neutron diffractometry in industrial and university settings. Such facilities will utilise a compact electron linear accelerator (linac) to deliver a \qty{35}{MeV}, kilowatt-scale electron beam to a TMR optimised to generate slow neutrons in the \qtyrange{1.5}{3.5}{\angstrom} wavelength range for time of flight measurements. A conceptual visualisation is shown in \autoref{fig:VULCAN_Schematic}, with further details provided in Refs.\cite{VULCANUpcoming, ElectronDriven}.

Key to the success of VULCAN is a TMR design that efficiently converts the electron beam into the desired slow neutrons. To validate the simulation-based design framework, a prototype TMR was constructed and characterised at CERN's CLEAR facility. This paper presents the prototype design, experimental setup, and initial results. 

The paper is structured as follows: \autoref{sec:ExperimentalOverview} provides an overview of the experiment, \autoref{sec:TMR_Proto} describes the prototype TMR design, and \autoref{sec:ExpSetup} details the experimental setup at CLEAR, including installation,  modelling, and data processing. \autoref{sec:HeMeasurements} compares the measured neutron yields and spectra with simulation predictions and discusses potential causes for observed discrepancies. \autoref{sec:Conclusions} summarises the findings and discusses next steps for future testing.

\section{Experimental Overview} \label{sec:ExperimentalOverview}

The VULCAN concept has been under development since 2021 by an industrial consortium of DAES SA, the Danish Technological Institute, and Xnovo Technology ApS, with support from CERN since 2023. To ensure that the compactness, cost, and complexity of the facility are compatible with its intended performance and business model, VULCAN operates as a time of flight neutron source driven by a pulsed electron linac~\cite{VULCANUpcoming}. 

Dedicated research and development efforts have focused on the design of a novel, compact TMR assembly capable of converting a pulsed \qty{35}{MeV} incident electron beam into the desired thermal neutron spectrum. The design targets production of \qtyrange{6}{36}{meV} neutrons (wavelengths \qtyrange{1.5}{3.5}{\angstrom}, velocities \qtyrange{1.1}{2.6}{km/s}) with an initial pulse full width at half maximum (FWHM) of $\leq$~\qty{20}{\micro s} to achieve the needed resolution within a sufficiently compact flight path (order of \qty{10}{m}) for industrial deployment.

The TMR design has been iteratively developed using the Monte Carlo radiation transport code \textsc{FLUKA}~\cite{FLUKA1,FLUKA2,FLAIR}. To enable early-stage validation of this design process, a prototype TMR assembly based on a decoupled, cold liquid methane moderator and without active target cooling was fabricated to compare measured neutron flux, energy spectrum, and pulse FWHM with simulation predictions. As modelling studies indicated that the temporal pulse structure has significant dependence on neutron poisoning, the prototype was designed to operate in both poisoned and unpoisoned configurations to further validate the modelling framework.

CERN's CLEAR (CERN Linear Electron Accelerator for Research) facility provided the electron beam for testing the TMR prototype~\cite{CLEARFacility}. Its \qty{30}{m}-long beamline includes a spectrometer for energy measurement and a comprehensive suite of diagnostics, including bunch charge monitors, beam position monitors, and YAG screens for beam characterisation. The accelerator optics, as well as the power and phase of the RF fields delivered to each of the four accelerating structures, can be individually and remotely set, enabling a wide range of beam parameters in energy, energy spread, charge, bunch length, and transverse size. CLEAR also provides short access times to the accelerator hall, with post-operation cooldowns of only \qtyrange{15}{30}{\minute} permitting multiple adjustments of the experimental setup throughout the day. 

\autoref{tab:CLEAR_Params} summarises the electron beam parameters required for the final VULCAN design, those used for optimising the prototype design in simulation, and those attainable at the CLEAR facility. As the prototype TMR lacks active cooling, it was designed to operate at substantially reduced beam power. The main discrepancies between the simulation-optimised conditions and those achievable at CLEAR were the beam energy and transverse beam size at the target. The beam could not be reliably transported below \qty{40}{MeV}, and its transverse size could not be reduced below \qty{4}{mm} owing to the relatively large distance between the final set of focusing quadrupoles in CLEAR and the TMR target. To enable direct comparison between simulation and experiment, the beam energy, size, and charge were recorded for each measurement and simulations were subsequently performed using these recorded parameters as inputs.

\begin{table}[htbp]
\centering
\caption{Electron beam parameters required for the final VULCAN design, used for optimising the prototype designs, and achievable at CERN’s CLEAR facility.}
\label{tab:CLEAR_Params}
    \begin{tabular}{@{}lccc@{}}
    \toprule
    \textbf{Parameter} & \textbf{VULCAN Facility} & \textbf{Prototype Optimisation} & \textbf{CLEAR Attainable} \\
    \midrule
    Beam energy & \qty{35}{MeV} & \qty{35}{MeV} & \qty{40}{MeV} \\
    Energy spread & $<$~\qty{1}{MeV} & $<$~\qty{1}{MeV} & $<$~\qty{1}{MeV} \\
    Beam size & \qty{0.5}{mm} & \qty{0.5}{mm} & \qty{4}{mm} \\
    Train length & $<$~\qty{1}{\micro\second} & $<$~\qty{1}{\micro\second} & $<$~\qty{100}{ns} \\
    Train charge & $\gtrsim$~\qty{290}{nC} & – & \qty{4}{nC} \\
    Repetition rate & $\sim$~\qty{100}{Hz} & – & \qty{10}{Hz} \\
    Beam power & $>$~\qty{1}{kW} & $>$~\qty{1}{W} & \qty{1.6}{W} \\
    \bottomrule
    \end{tabular}
\end{table}

\section{TMR prototype} \label{sec:TMR_Proto}

\begin{figure}[tbh!]
     \centering
     \begin{subfigure}[b]{0.32\textwidth}
         \centering
         \includegraphics[width=1\textwidth]{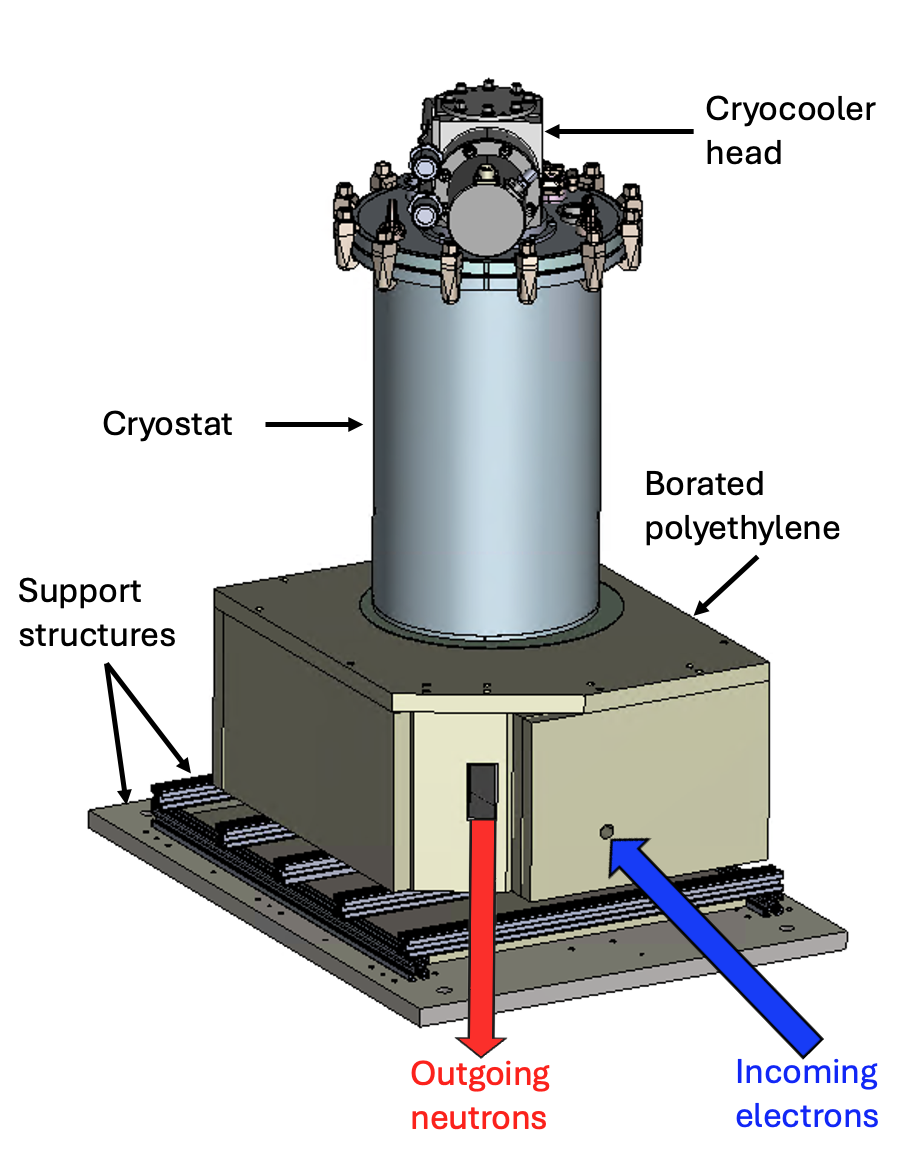}
         \caption{External view.}
         \label{fig:TMR_1}
     \end{subfigure}
     \begin{subfigure}[b]{0.32\textwidth}
         \centering
         \includegraphics[width=1\textwidth]{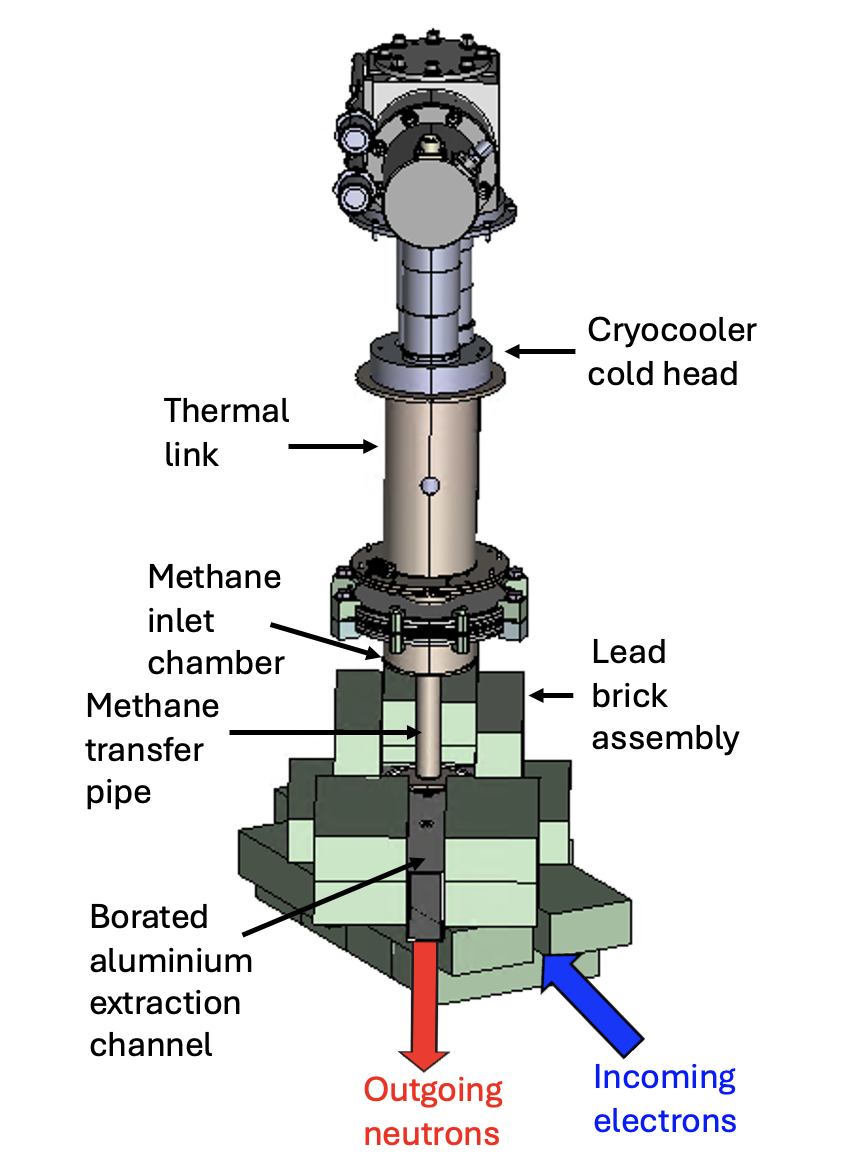}
         \caption{Internal cryostat view.}
         \label{fig:TMR_2}
     \end{subfigure}
     \begin{subfigure}[b]{0.32\textwidth}
         \centering
         \includegraphics[width=1\textwidth]{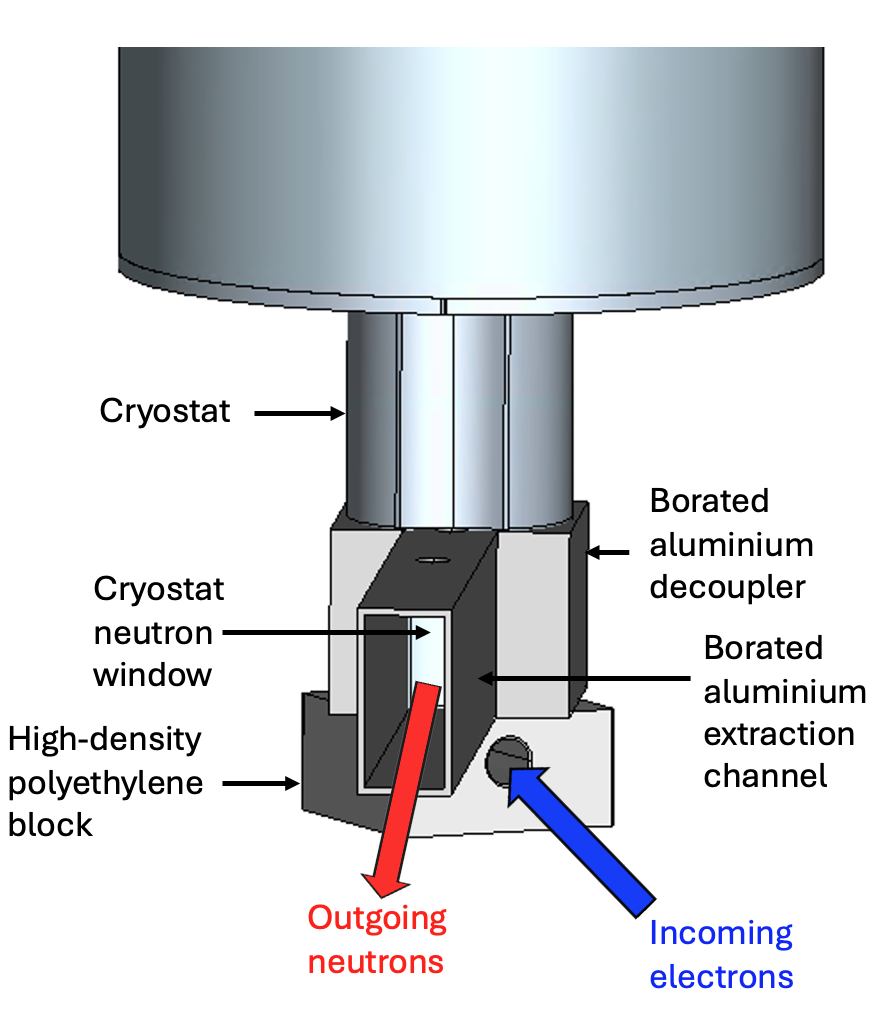}
         \caption{Decoupler view.}
         \label{fig:TMR_3}
     \end{subfigure}
     \begin{subfigure}[b]{0.32\textwidth}
         \centering
         \includegraphics[width=1\textwidth]{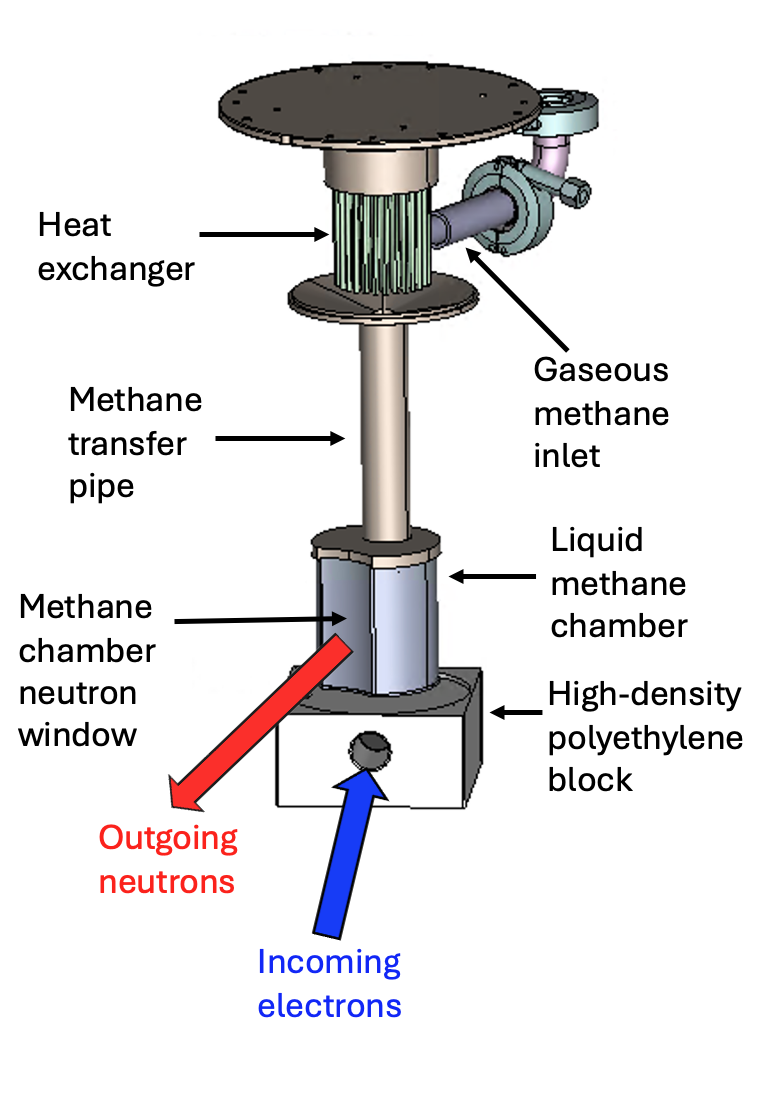}
         \caption{Methane chamber view.}
         \label{fig:TMR_4}
     \end{subfigure}
     \begin{subfigure}[b]{0.32\textwidth}
         \centering
         \includegraphics[width=1\textwidth]{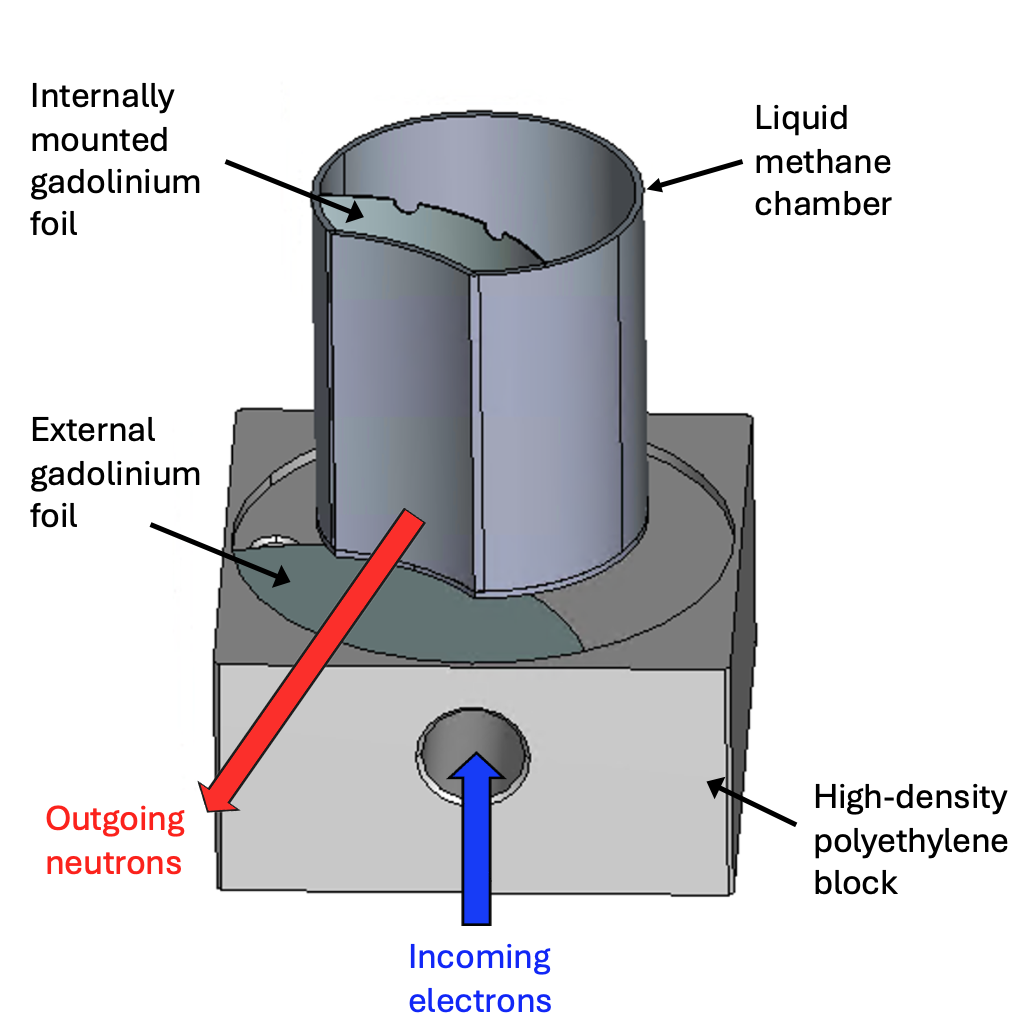}
         \caption{Poison view.}
         \label{fig:TMR_5}
     \end{subfigure}
     \begin{subfigure}[b]{0.32\textwidth}
         \centering
         \includegraphics[width=1\textwidth]{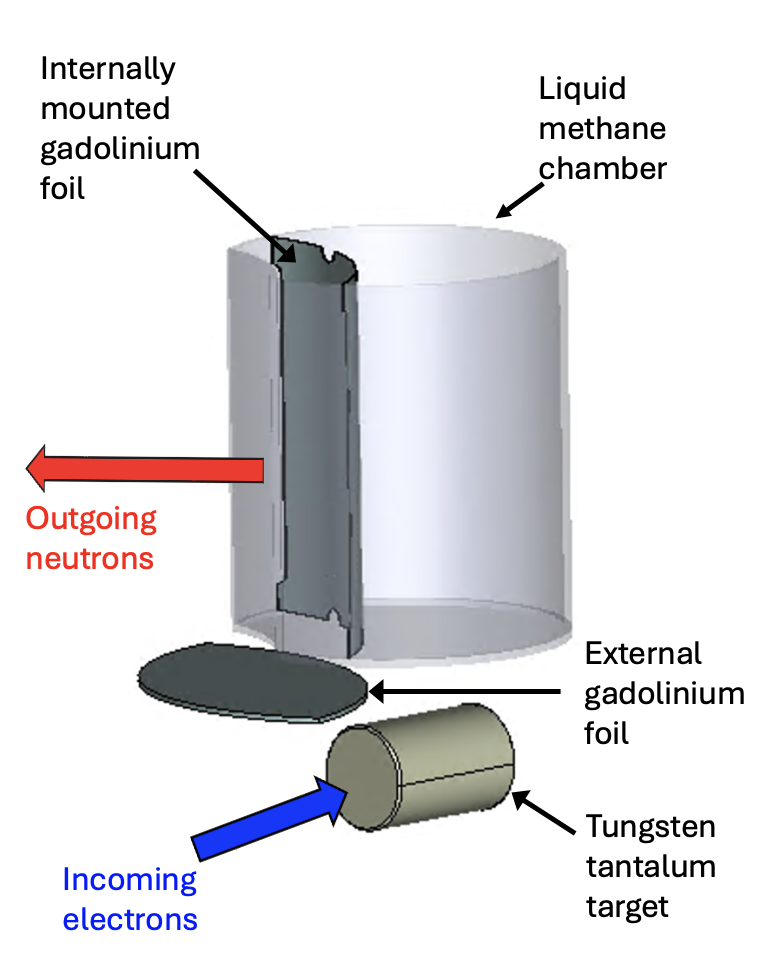}
         \caption{Target view.}
         \label{fig:TMR_6}
     \end{subfigure}
        \caption{The decoupled-poisoned TMR prototype.}
        \label{fig:TMR_CAD}
\end{figure}

\autoref{fig:TMR_CAD} shows the design of the prototype TMR in the poisoned configuration. \autoref{fig:TMR_1} presents an external view of the fully assembled structure, \autoref{fig:TMR_2} shows the interior of the cryostat with the cryostat and borated polyethylene shielding removed, while \autoref{fig:TMR_3} highlights the borated aluminium decoupler with the outer shielding and lead brick assembly removed. \autoref{fig:TMR_4} focuses on the methane chamber and its surrounding components, and \autoref{fig:TMR_5} and \autoref{fig:TMR_6} illustrate the spatial arrangement of the high-density polyethylene (HDPE) pre-moderator, methane moderator,  gadolinium poison, and tungsten tantalum target.

\begin{figure}[tbh!]
     \centering
     \begin{subfigure}[b]{0.49\textwidth}
         \centering
         \includegraphics[width=1\textwidth]{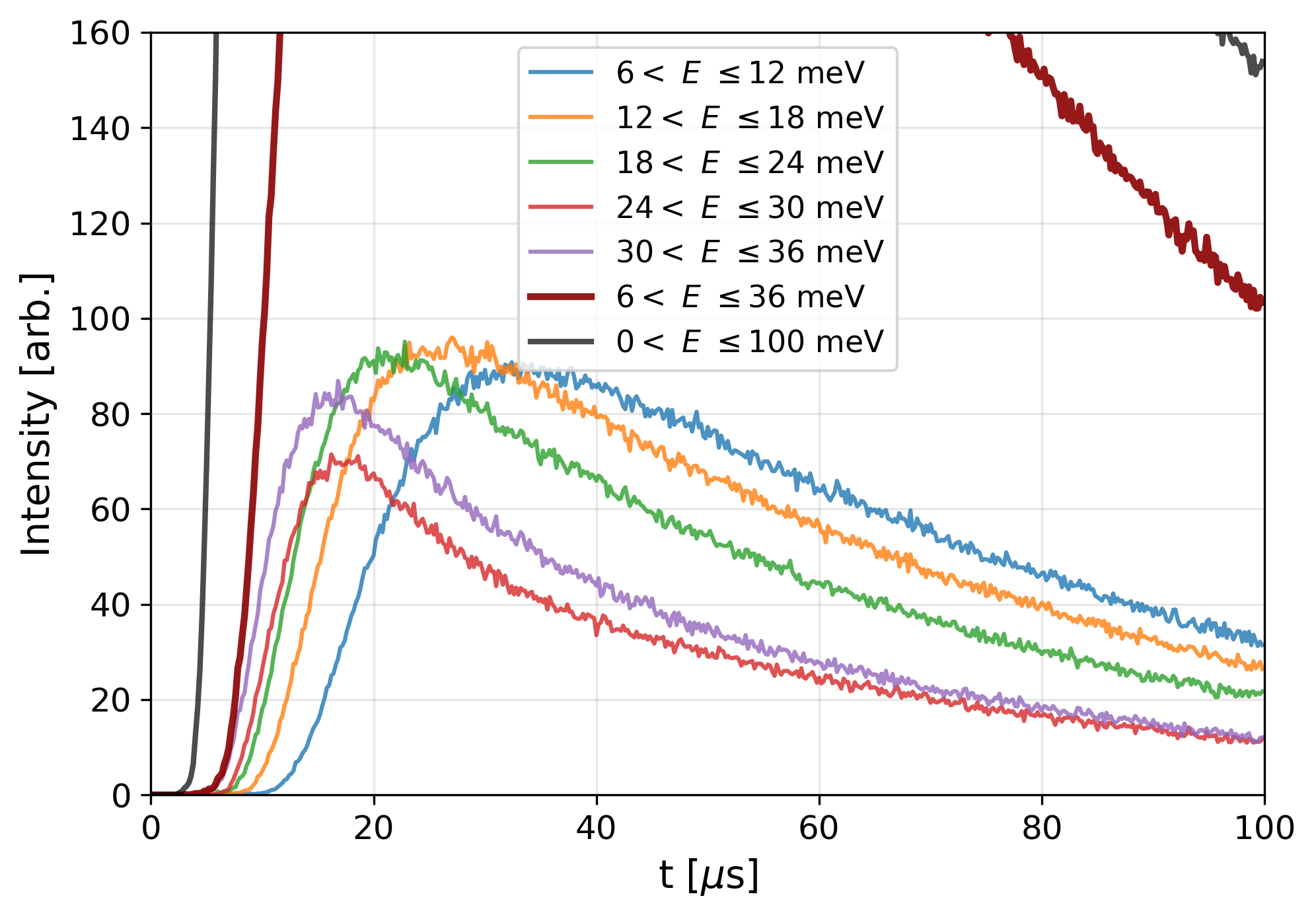}
         \caption{Unpoisoned.}
         \label{fig:UnpoisonedTime}
     \end{subfigure}
     \begin{subfigure}[b]{0.49\textwidth}
         \centering
         \includegraphics[width=1\textwidth]{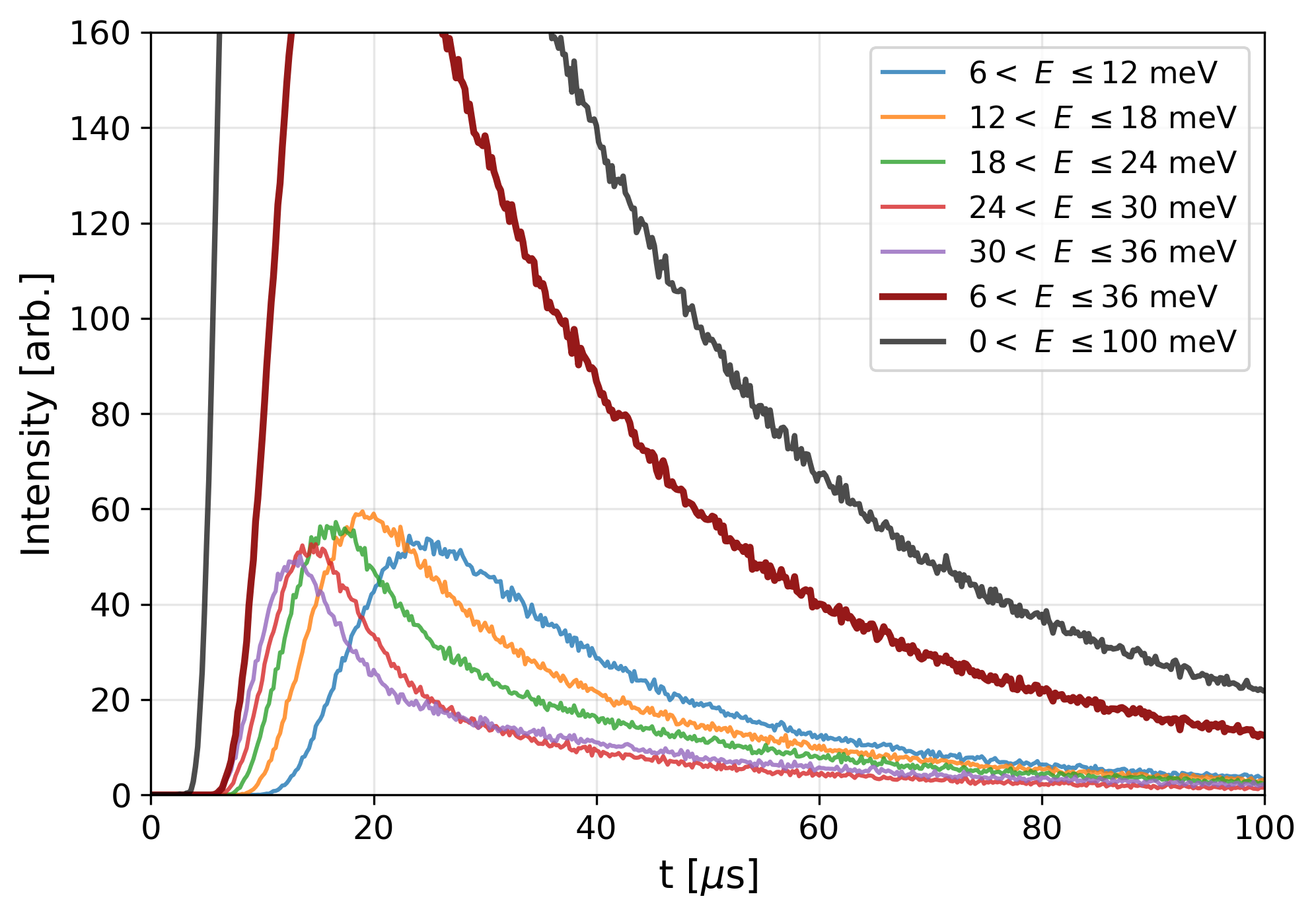}
         \caption{Poisoned.}
         \label{fig:PoisonedTime}
     \end{subfigure}
        \caption{Comparison between the time spectra of slow neutrons at the cryostat neutron window between the unpoisoned and poisoned configurations.}
        \label{fig:TimeSpectrum}
\end{figure}

The key, functional components of the TMR annotated in \autoref{fig:TMR_CAD} are as follows:
\begin{itemize}
    \item \textbf{Target:} The incident electron beam strikes a cylindrical tungsten tantalum (W10Ta90) target (\autoref{fig:TMR_6}), \qty{1}{cm} in radius and \qty{3}{cm} in length. High-energy (\qty{}{MeV}-scale peak) photons  are generated via the Bremsstrahlung process 
    \begin{align*}
        & {e^- + N \rightarrow e^- + N + \gamma},
    \end{align*}
    where $N$ represents a nucleus in the target. These photons subsequently produce photoneutrons (\qty{}{MeV}-scale peak) through the reaction
    \begin{align*}
        & {\gamma + N \rightarrow N^* \rightarrow N' + n},
    \end{align*}
    where $N^*$  denotes an excited nucleus. 

    \item \textbf{Pre-moderator:} The target is encased inside of a \qtyproduct{100x50x100}{mm} high-density polyethylene block (\autoref{fig:TMR_3} to \autoref{fig:TMR_5}), which serves the purpose of pre-moderating the photoneutron spectrum before it enters the cold, liquid methane moderator.
    
    \item \textbf{Moderator:} The methane chamber (\autoref{fig:TMR_4} to \autoref{fig:TMR_6}) is positioned inside the TMR cryostat with its bottom face \qty{15}{mm} above the HDPE pre-moderator. The chamber is cylindrical with a height of \qty{70}{mm} and a circular cross-section of radius \qty{30}{mm}, except for the removal of a convex minor segment to create a concave neutron window surface. In operation, the methane chamber holds approximately \qty{85}{g} (\qty{200}{mL}) of liquid methane at \qty{100}{K} to moderate neutrons to the \qty{}{meV} range. The methane chamber is cooled by a \qty{2}{}-stage cryocooler (\autoref{fig:TMR_1} and \autoref{fig:TMR_2}).

    \item \textbf{Poison:} Two different, interchangeable methane chambers were manufactured, allowing the TMR to be configured in either a decoupled-poisoned or decoupled-unpoisoned setup.
        \begin{itemize}
            \item \textbf{Poisoned:} The poisoned configuration includes a \qty{0.05}{mm}-thick gadolinium foil internally mounted within the methane chamber (\autoref{fig:TMR_5} and \autoref{fig:TMR_6}) and an additional external gadolinium foil placed beneath the chamber between the HDPE pre-moderator and the cryostat (\autoref{fig:TMR_5} and \autoref{fig:TMR_6}). The internal foil is curved with the same radius of curvature as the neutron window, maintaining a separation of \qty{16}{mm} between the foil and window surface, optimised for neutron flux and temporal pulse structure\footnote{Future TMR designs will also optimise for resolution by, for example, incorporating a flatter moderator surface.}.
            
            \item \textbf{Unpoisoned:} The unpoisoned configuration uses a methane chamber without any internal poison and omits the external gadolinium foil beneath the cryostat.
        \end{itemize}  
    The effect of neutron poisoning on the temporal pulse structure is illustrated in \autoref{fig:TimeSpectrum}, which shows neutron time spectra at the cryostat neutron window for both configurations. The poisoned configuration approximately reduces both the flux and the initial pulse FWHM of thermal neutrons in the \qtyrange{6}{36}{meV} range by a factor of two.
    
    \item \textbf{Decoupler:} A \qty{3}{mm}-thick borated aluminium (0.31 mass fraction boron carbide, 0.69 mass fraction Al-6061 alloy) shell (\autoref{fig:TMR_3}) surrounds the cryostat volume housing the methane chamber. The shell is positioned between the cryostat and the lead reflector and provides neutron absorption on all surfaces except the bottom face and the neutron extraction channel, decoupling the moderator from the surrounding reflector.
    
    \item \textbf{Reflector:} An assembly of lead bricks (\autoref{fig:TMR_2}) surrounds the HDPE pre-moderator and the methane chamber.
    
    \item \textbf{Shielding}: The target, pre-moderator, moderator, and reflector components are all enclosed within a \qty{25}{mm}-thick \qty{5}{\%} borated polyethylene box (\autoref{fig:TMR_1}). 
    
    \item \textbf{Neutron extraction channel:} A \qtyproduct{40x80}{mm} borated aluminium extraction channel (\autoref{fig:TMR_2} and \autoref{fig:TMR_3}) links the concave surface of the cryostat to the outside of the borated polyethylene box. The methane chamber neutron window, cryostat neutron window, and extraction channel are offset from the incoming electron beam direction by \ang{30}.
    
    \item \textbf{Cryostat:} The 316L stainless steel cryostat (\autoref{fig:TMR_1} and \autoref{fig:TMR_2}) encloses the cryocooler head and methane chamber, and is evacuated to provide thermal insulation and prevent moisture in the air from freezing on the cryogenic components. The lower section of the cryostat housing the methane chamber is machined with a cryostat neutron window (\autoref{fig:TMR_3}), a concave surface aligned with the methane chamber neutron window. A gap of \qty{12.5}{mm} separates the methane chamber window from the cryostat window.
    
    \item \textbf{Gas line:} A gas line (not shown in \autoref{fig:TMR_CAD}) is used to flush and evacuate the methane chamber prior to safely filling it with methane. After operation, the gas line is also used to safely evacuate the methane from the TMR. 

\end{itemize}

\section{Experiment at CLEAR} \label{sec:ExpSetup}

\subsection{Installation}

\subsubsection{TMR}

The cryostat and both poisoned and unpoisoned methane chambers were commissioned in CERN’s Cryolab. Leak testing verified a leak rate below \qty{1e-7}{mbar\cdot l/s} throughout the system and vacuum tests showed the required vacuum level and stability were attained. The complete operational procedure of gas injection into the methane chamber with a mass flow controller, liquification, storage, and evacuation was also confirmed with argon as a substitute for methane, as the necessary safety measures were not in place for handling methane in the Cryolab. A PID-based temperature control system demonstrated the ability to maintain the methane chamber temperature within $\pm$\qty{2}{K}.

\begin{figure}[tbh!]
     \centering
     \begin{subfigure}[b]{0.32\textwidth}
         \centering
         \includegraphics[width=1\textwidth]{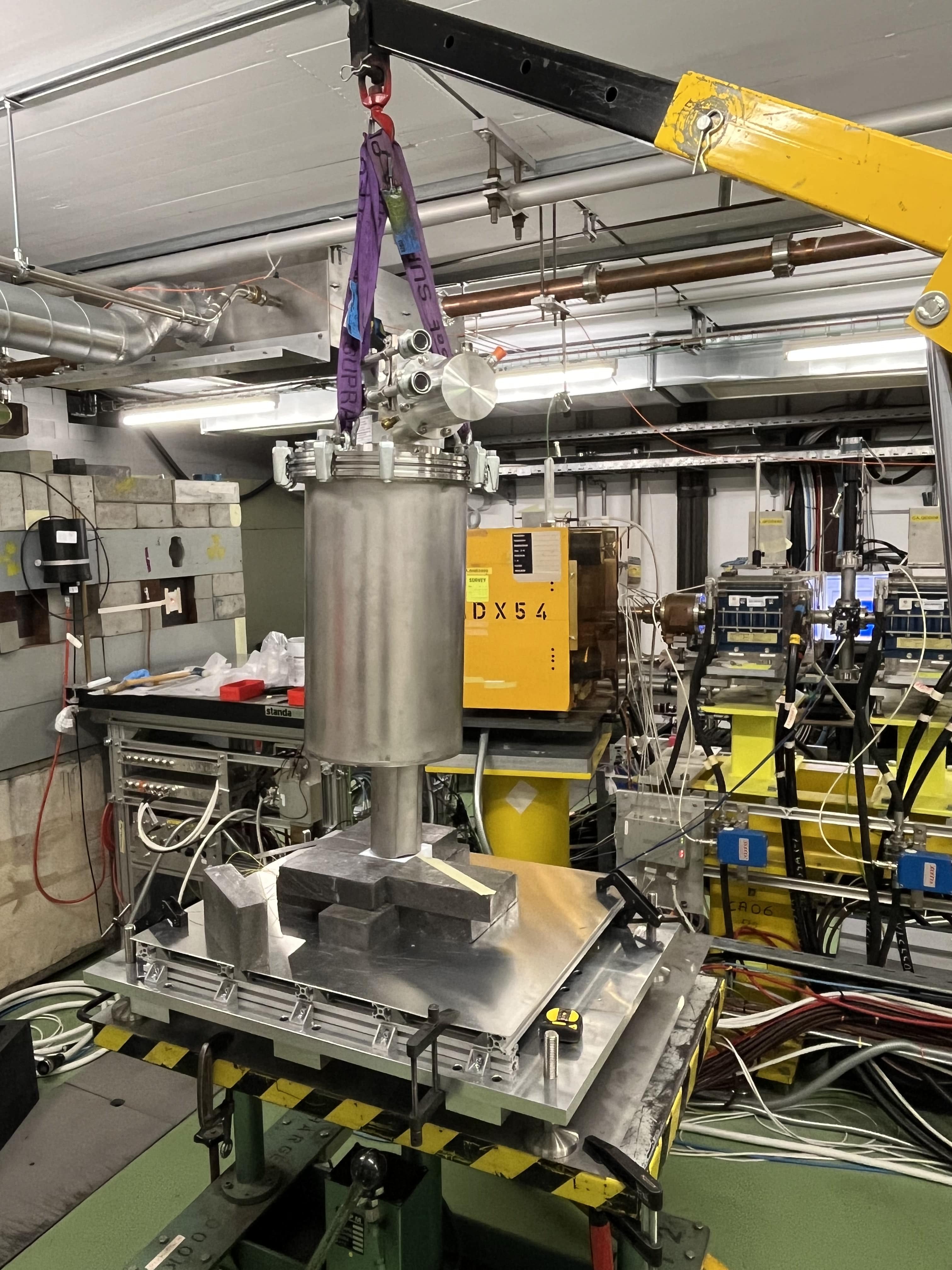}
         \caption{Assembly.}
         \label{fig:CLEAR_1}
     \end{subfigure}
     \begin{subfigure}[b]{0.32\textwidth}
         \centering
         \includegraphics[width=1\textwidth]{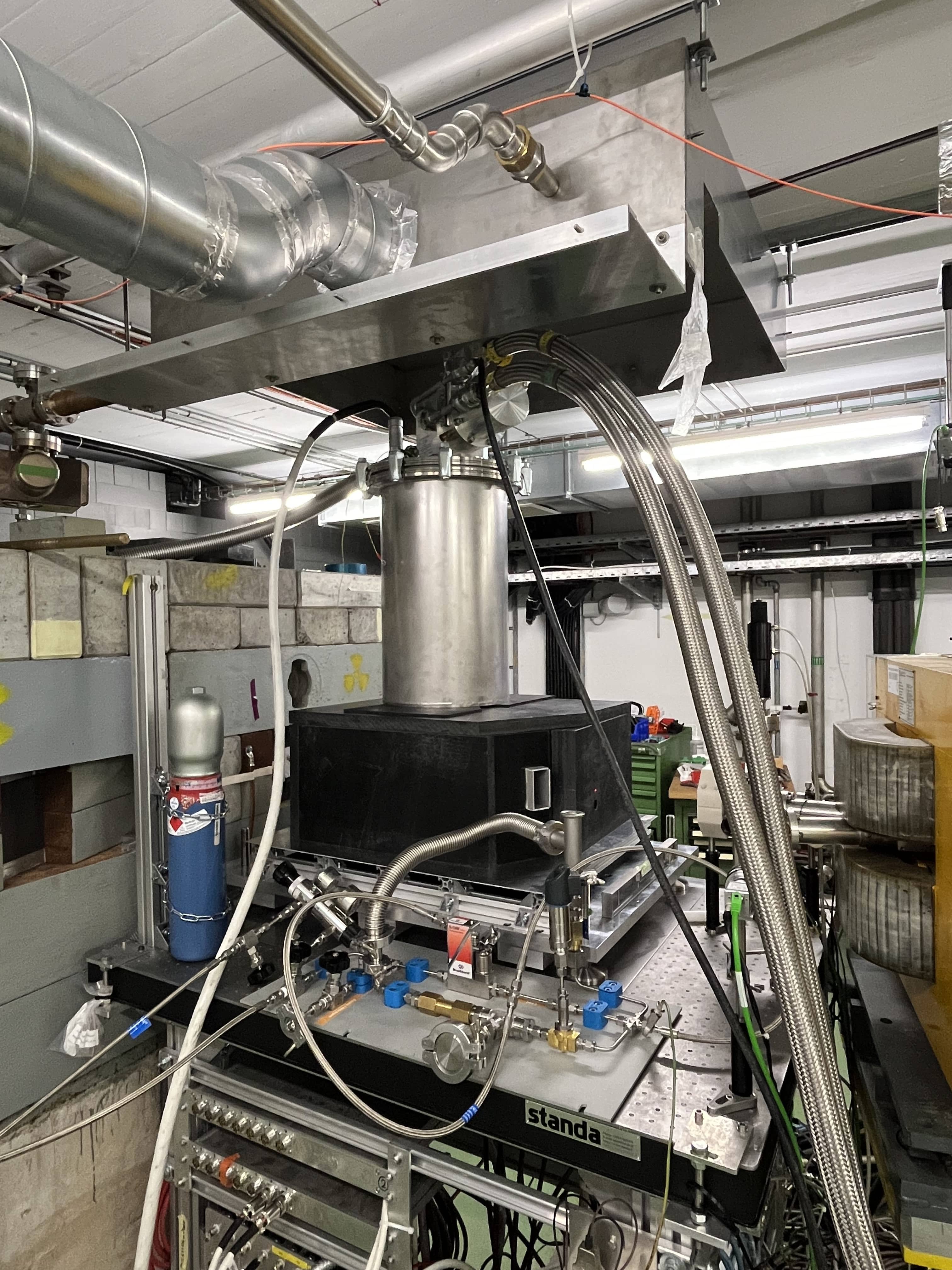}
         \caption{Installation.}
         \label{fig:CLEAR_2}
     \end{subfigure}
        \begin{subfigure}[b]{0.32\textwidth}
         \centering
         \includegraphics[width=1\textwidth]{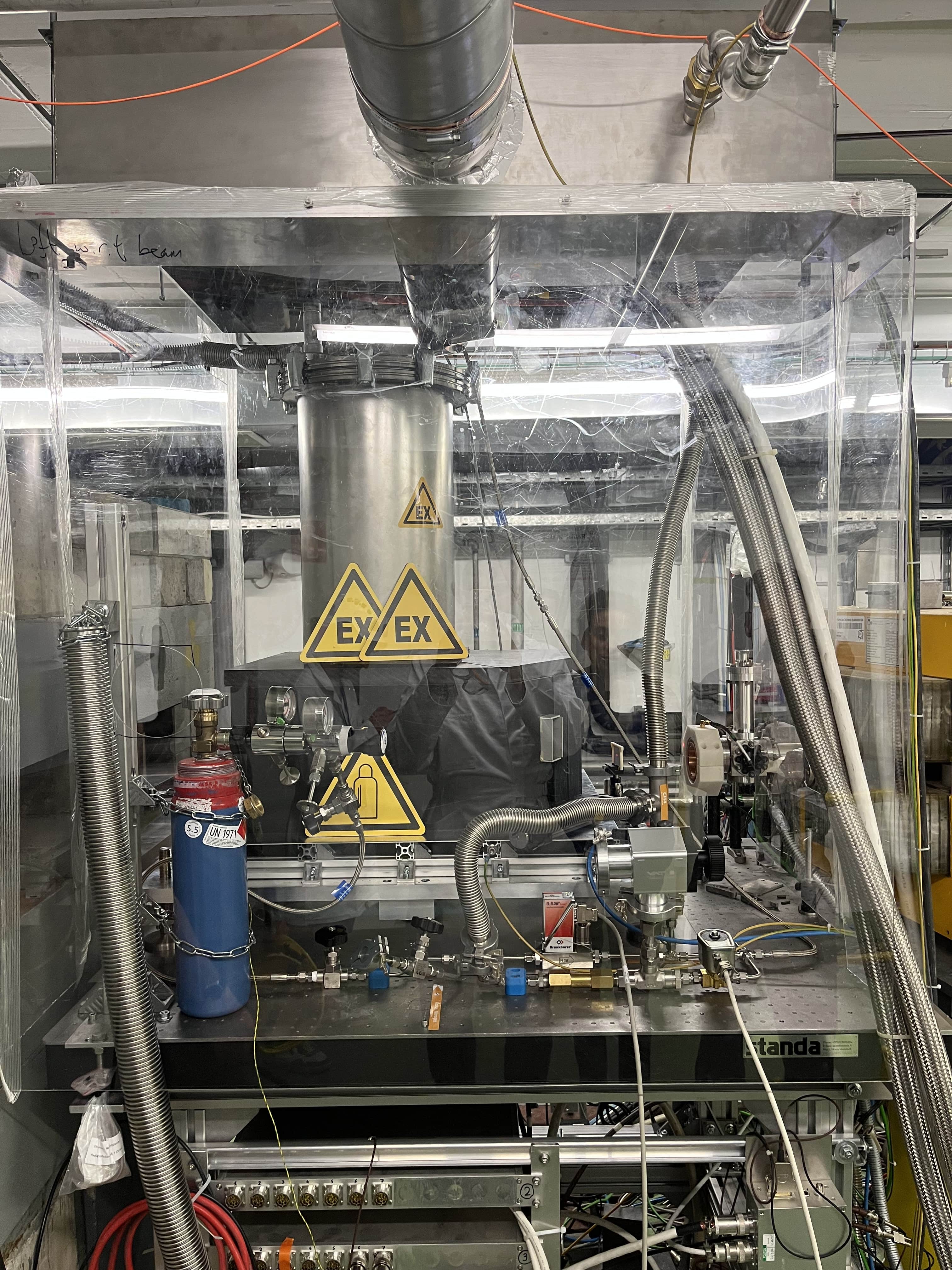}
         \caption{Ready for filling.}
         \label{fig:CLEAR_3}
     \end{subfigure}
        \caption{The assembly and installation of the VULCAN TMR in CLEAR.}
        \label{fig:CLEAR_Installation}
\end{figure}

The TMR was then fully assembled and installed on CLEAR's in-air test stand, with \autoref{fig:CLEAR_Installation} illustrating the various stages. Following installation, the centre of the TMR target was aligned with the beam axis using an alignment laser mounted in the CLEAR beamline. A LANEX scintillating screen was affixed to the target face to enable direct measurement of the beam position and size on the target. A ventilation hood and Makrolon curtain were also installed around the test stand to contain any potential methane leakage. 

\subsubsection{Detector}

\begin{figure}[tbh!]
     \centering
     \begin{subfigure}[b]{0.281\textwidth}
         \centering
         \includegraphics[width=1\textwidth]{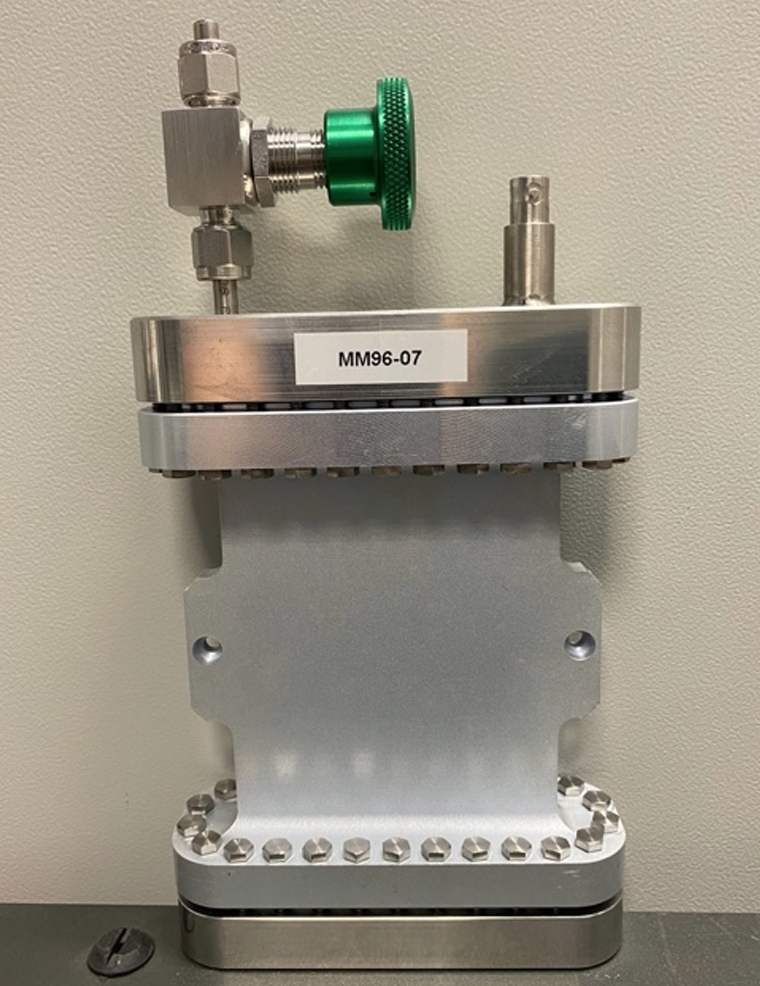}
         \caption{$^3$He detector~\protect\cite{ILL_Detector}.}
         \label{fig:He_1}
     \end{subfigure}
     \begin{subfigure}[b]{0.65\textwidth}
         \centering
         \includegraphics[width=1\textwidth]{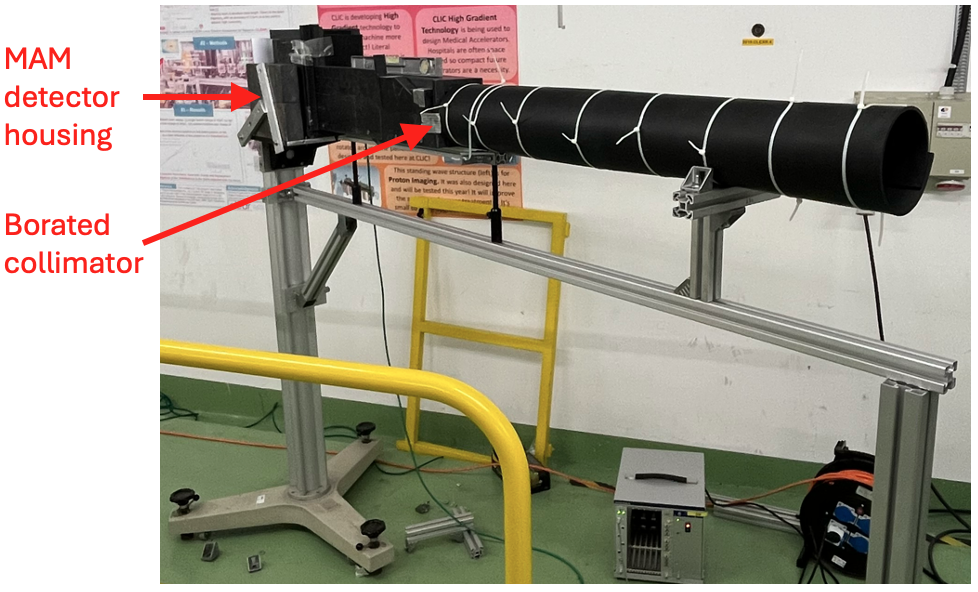}
         \caption{Installation in CLEAR.}
         \label{fig:He_2}
     \end{subfigure}
        \caption{The $^3$He detector in the experimental campaign.}
        \label{fig:He_detectors}
\end{figure}

A $^3$He Monoblock Aluminium Multitube (MAM) detector~\cite{ILL_Detector} was used to measure the thermal neutron spectra. The detector, shown in \autoref{fig:He_1}, consists of eight gas-filled proportional counter tubes, with a total active area of \qtyproduct{76x96}{mm}. It was installed at distances of \qty{1.6}{m}, \qty{4.2}{m}, and \qty{6.6}{m} from the cryostat neutron window and, to reduce background noise from the prompt $\gamma$ flash and from radiation originating outside the TMR, the detector was housed within shielding composed of lead and borated polyethylene as shown in \autoref{fig:He_1}. Additional collimation was provided by a tunnel of borated rubber, to further suppress off-axis radiation and improve signal-to-noise in the measured neutron spectrum.

Each tube of the detector was pressurised with \qty{1}{bar} of $^3$He, chosen to optimise the neutron detection efficiency\footnote{The peak measured instantaneous count rate was \qty{60000}{}~cps per tube. With the detector's intrinsic dead-time of $<$~\qty{1}{\micro s}, the counting saturation loss is $<$~\qty{6}{\%}.}. The detector was operated at a voltage of \qty{1.1}{kV} to: maximise detection efficiency without over-saturation, allow discrimination between high-energy neutron interactions and lower-energy gamma interactions, and protect the detector from potential overvoltage damage. The detector was configured with \qty{10}{\micro s} bin widths and its detection threshold was set above the electronic noise floor. A \qty{5}{\micro s} acquisition delay was applied after each beam pulse to suppress the prompt radiation background\footnote{The thermal neutrons arrive well-after the prompt radiation; for example, at the distance of \qty{1.6}{m}, \qty{25}{meV} thermal neutrons arrive at approximately \qty{700}{\micro s} after the prompt radiation.}.

\subsection{Modelling} \label{sec:ModSetup}

\begin{figure}
    \centering
    \includegraphics[width=0.75\linewidth]{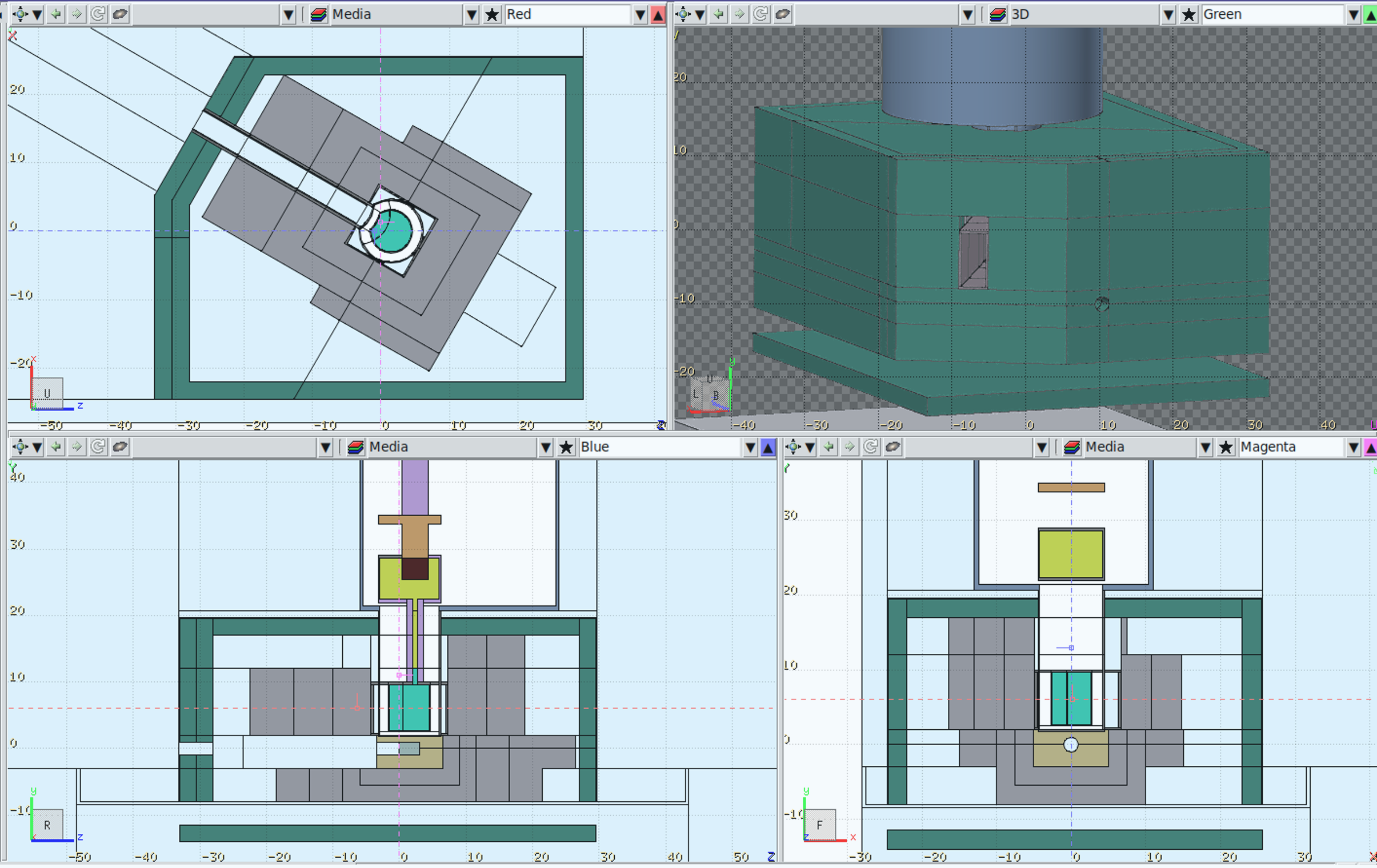}
    \caption{The prototype TMR assembly, as modelled in \textsc{FLUKA}.}
    \label{fig:FLUKA_Setup}
\end{figure}

The experimental setup in CLEAR was modelled using the \textsc{FLUKA4-5.0} Monte Carlo code, as shown in \autoref{fig:FLUKA_Setup}. Application of appropriate variance reduction techniques allowed modelling of the electron beam interaction with the target, photoneutron production, and neutron moderation processes in a single-step calculation. The pointwise low-energy neutron treatment was activated with the JEFF-3.3 neutron cross-section library applied to all materials. The S$(\alpha,\beta)$ molecular binding correction and Doppler broadening corresponding to liquid methane at \qty{100}{K} were applied to the moderator material. All other cold materials were modelled at \qty{100}{K}; except for aluminium and iron which were modelled at \qty{80}{K} due to the unavailability of S$(\alpha,\beta)$ data at \qty{100}{K}. 

The \textit{USRBDX} card was used to score the average differential flux of neutrons on \qty{73.0}{cm^2} surfaces placed at the same locations as the detector in the experiments. The incident electron beam was input with beam parameters recorded for each individual experiment.

\subsection{Data Processing} \label{sec:Processing}

Counts were recorded with the $^3$He detector over periods of one hour or longer, with the beam charge and size monitored and adjusted as necessary to maintain the beam specification listed in \autoref{tab:CLEAR_Params}.  To compare the neutron flux and spectra between the experiments and replicating simulations, data was processed into a differential flux $d\phi/dE$.

\subsubsection{Detector Data} \label{sec:HeProcessing}

The $^3$He detector resolved individual counts into one of \qty{256}{} energy channels and one of \qty{2000}{} time-channels. The time of flight principle enables conversion of the time channel data into an estimate of neutron energy distribution as
\begin{equation} \label{eqn:EnergyConversion}
    E = \frac{1}{2}m_0\left(\frac{L}{t}\right)^2,
\end{equation}
where $t$ is the neutron arrival time, $L$ is the detector-to-source-distance, and $m_0$~=~\qty{1.675e-27}{kg} is the mass of the neutron\footnote{The initial pulse width can be ignored because the shortest flight path of \qty{1.6}{m} is sufficiently long (i.e. at \qty{1.6}{m}, the \qty{700}{\micro s} flight time of a \qty{25}{meV} neutron is much greater than the \qty{20}{\micro s} and \qty{60}{\micro s} pulse widths of the poisoned and unpoisoned TMRs).}. 

The differential neutron flux is then calculated as
\begin{equation} \label{eq:ThermalDiffNFlux}
    \frac{d\phi}{dE}(E_i) = \frac{N_T(E_i)}{A_{\text{det}} E_{\text{target}} \eta(E_i) \Delta E_i},
\end{equation}
where $N_T(E_i)$ is the number of counts in the energy bin centered at $E_i$ with width $\Delta E_i$, $A_\text{det} = \qty{73}{cm^{2}}$ is the active detector area, $E_\text{target}$ is the total beam energy delivered to the target during a given measurement, and $\eta(E_i)$ is the energy-dependent detection efficiency (for example, \qty{10}{\%} at \qty{1.8}{\angstrom}).

\subsubsection{Processing Simulation Data}

\textsc{FLUKA} outputs the differential flux $F(E_i)$ in units of [\qty{}{GeV^{-1}cm^{-2}pr^{-1}}], where \qty{}{pr} denotes the number of primary electrons. The corresponding differential flux, consistent with \autoref{eq:ThermalDiffNFlux}, is therefore
\begin{equation}~\label{eq:ThermalDiffNFluxFLUKA}
    d\phi/dE(E_i) = F(E_i)N_e,
\end{equation}
where $N_e=\qty{1.5e14}{}~[\qty{}{pr/s/kW}]$ is the number of primary electrons per second per kilowatt for a \qty{40}{MeV} beam.

\section{Spectra Measurements and Comparison with Simulation}~\label{sec:HeMeasurements}

\subsection{Background}~\label{sec:BoratedGate}

\begin{figure}[tbh!]
     \centering
     \begin{subfigure}[b]{0.3\textwidth}
         \centering
         \includegraphics[width=1\textwidth]{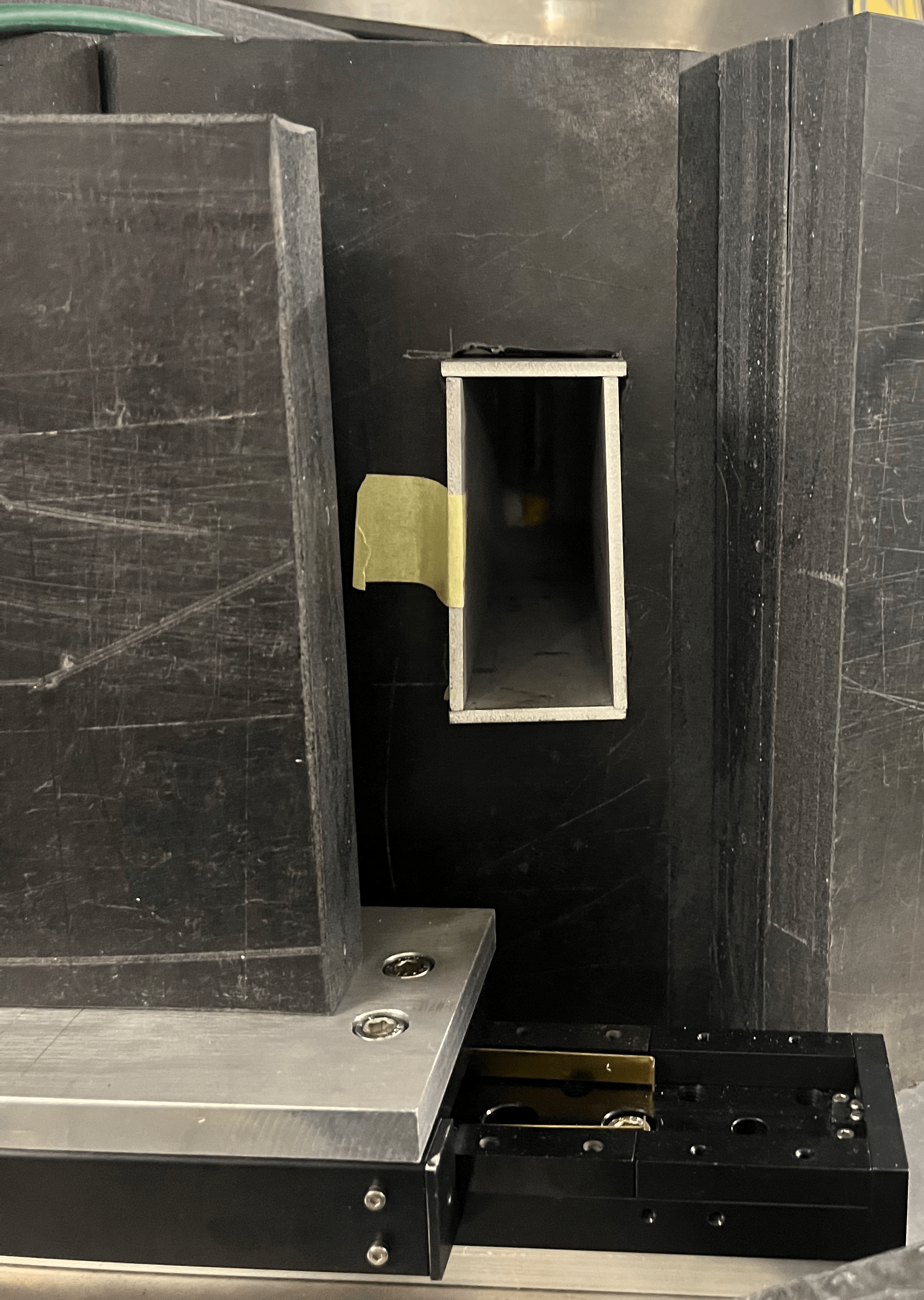}
         \caption{Borated gate.}
         \label{fig:BoratedGate}
     \end{subfigure}
     \begin{subfigure}[b]{0.49\textwidth}
         \centering
         \includegraphics[width=1\textwidth]{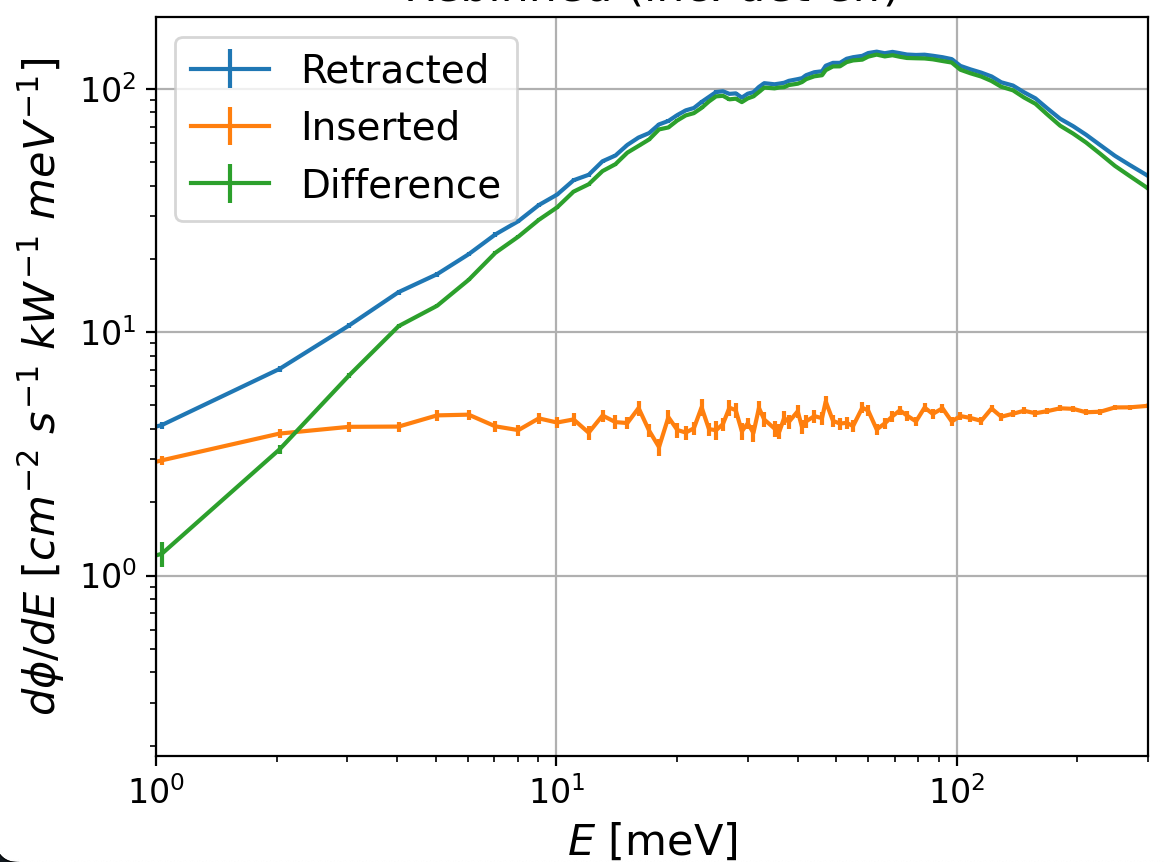}
         \caption{Spectrum.}
         \label{fig:BackgroundSpectrum}
     \end{subfigure}
        \caption{The borated gate shown in the extracted position, leaving the borated aluminium extraction channel exit unobstructed (left). The corresponding measured thermal neutron spectra are shown (right) for: the gate retracted, the gate inserted, and the difference between the two.}
        \label{fig:Background}
\end{figure}

To quantify the background signal in the $^3$He detector, a \qty{5}{cm}-thick borated polyethylene gate was mounted on a movable stage and installed near the aluminium extraction channel exit. The gate could either block the exit or be retracted from it, as shown in \autoref{fig:BoratedGate}.

\autoref{fig:BackgroundSpectrum} compares the thermal neutron spectrum measured at the distance of \qty{1.6}{m} from the cryostat neutron window with the borated gate inserted, retracted, and the difference between the two. Inserting the gate reduced the measured count rate by approximately \qty{95}{\%}, indicating that the majority of detected events originate from neutrons emitted through the aluminium extraction channel. 

\subsection{Spectra Comparison}~\label{sec:SpectraDistances}

\begin{figure}[tbh!]
     \centering
     \begin{subfigure}[b]{0.49\textwidth}
         \centering
         \includegraphics[width=1\textwidth]{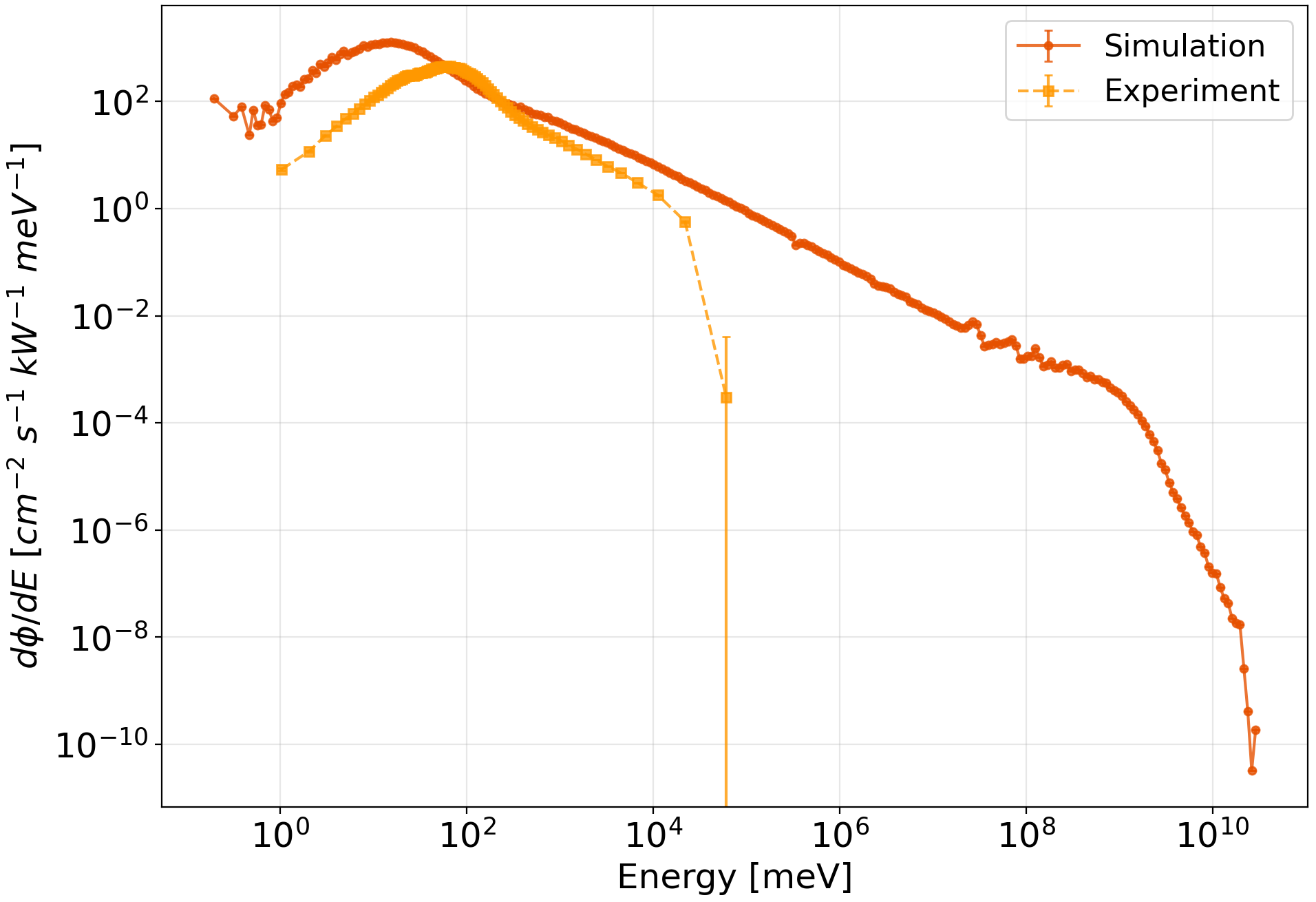}
         \caption{Unpoisoned.}
         \label{fig:Unpoisoned}
     \end{subfigure}
     \begin{subfigure}[b]{0.49\textwidth}
         \centering
         \includegraphics[width=1\textwidth]{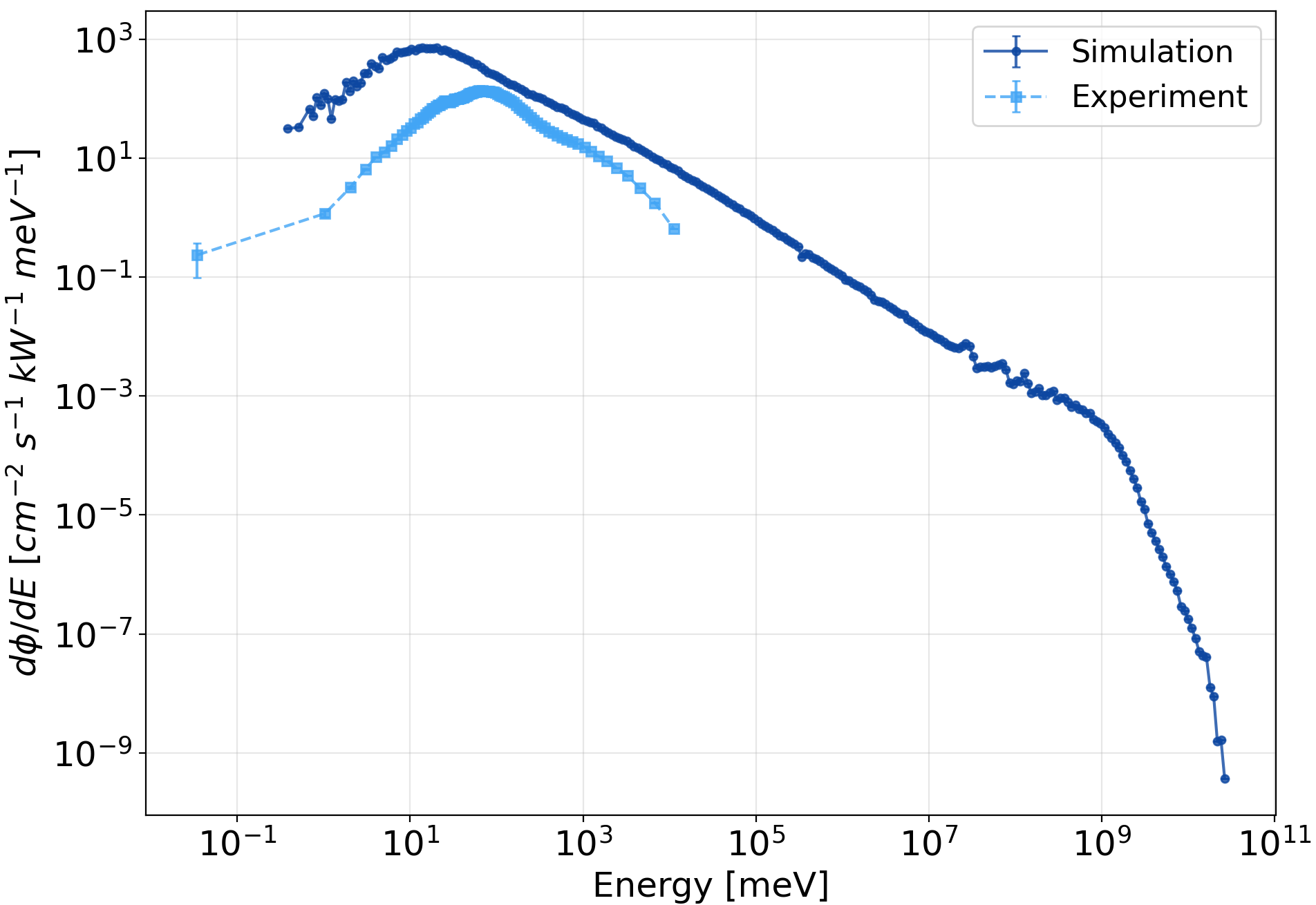}
         \caption{Poisoned.}
         \label{fig:Poisoned}
     \end{subfigure}
     \begin{subfigure}[b]{0.49\textwidth}
         \centering
         \includegraphics[width=1\textwidth]{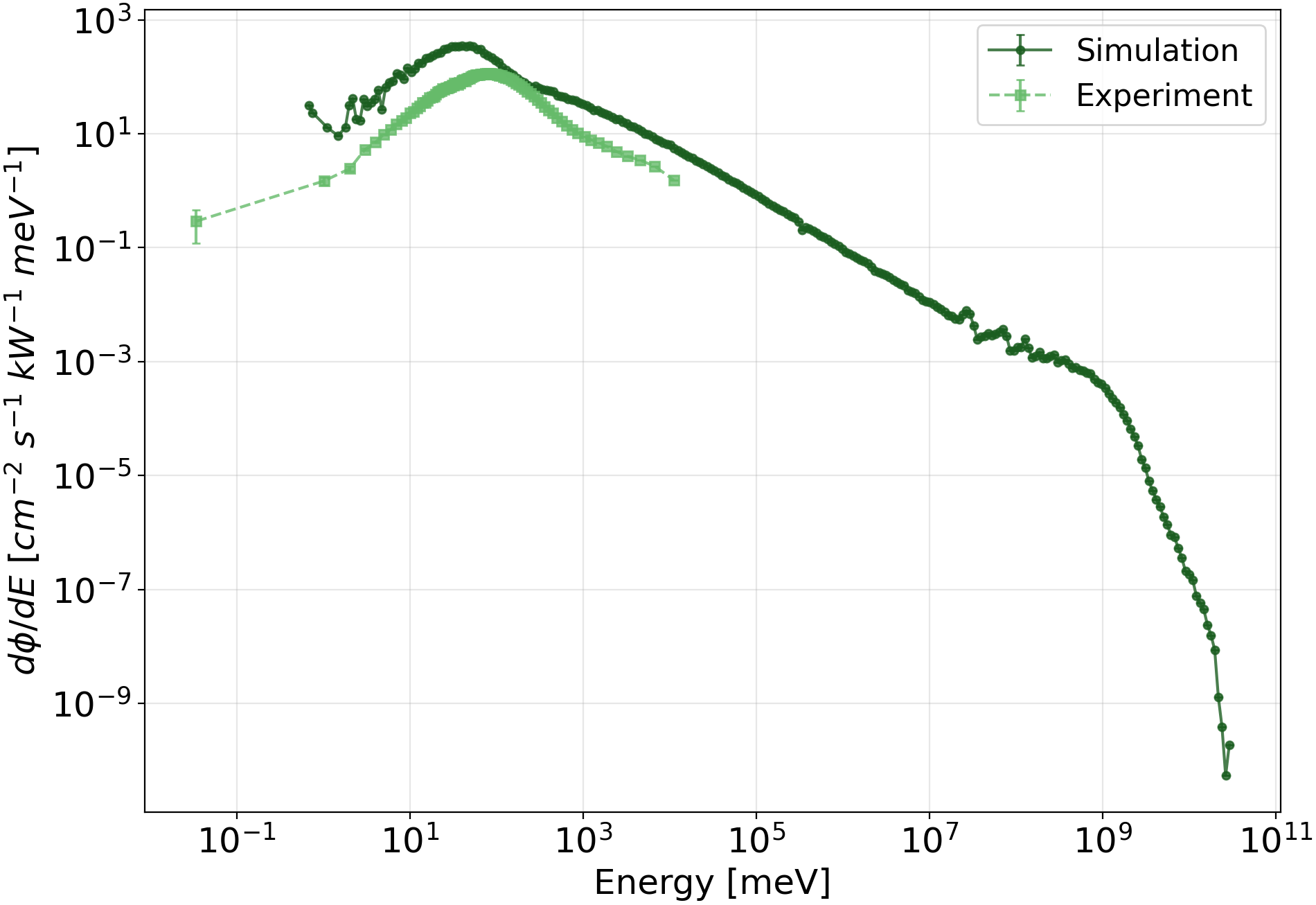}
         \caption{Unpoisoned vented.}
         \label{fig:Unpoisoned_Empty}
     \end{subfigure}
     \begin{subfigure}[b]{0.49\textwidth}
         \centering
         \includegraphics[width=1\textwidth]{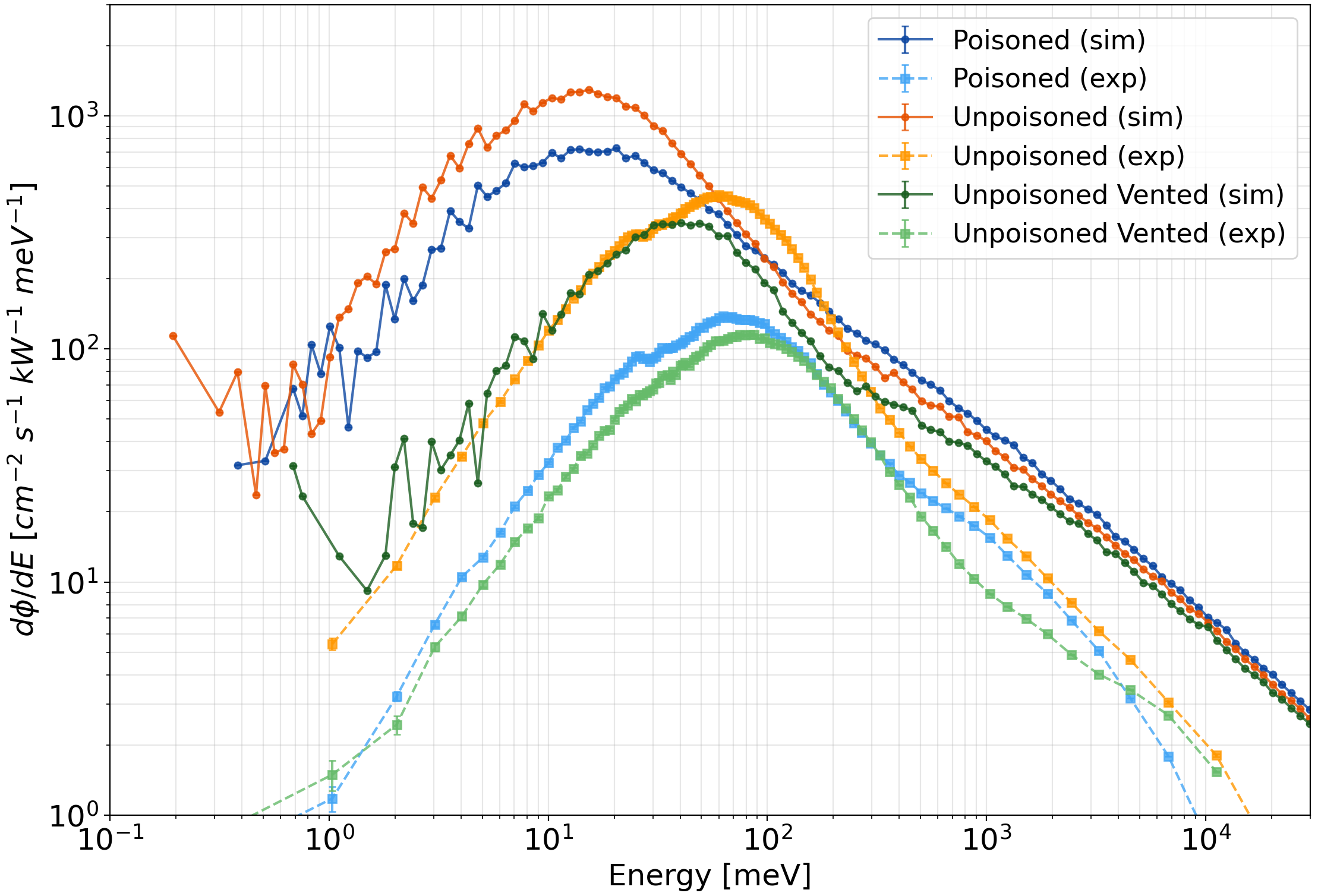}
         \caption{All.}
         \label{fig:All}
     \end{subfigure}
        \caption{Neutron spectra measured at \qty{1.6}{m} for the different experimental setups.}
        \label{fig:Spectra}
\end{figure}

\autoref{fig:Spectra} compares the neutron spectra at \qty{1.6}{m} for: the decoupled-unpoisoned configuration (\autoref{fig:Unpoisoned}), decoupled-poisoned configuration (\autoref{fig:Poisoned}), decoupled-unpoisoned configuration with the methane chamber vented (\autoref{fig:Unpoisoned_Empty}), and with all datasets on the same plot (\autoref{fig:All}). 

The most notable observation is the clear discrepancy in peak energy between experiment (\qty{65}{meV}) and simulation (\qty{15}{meV}) for both unpoisoned (\autoref{fig:Unpoisoned}) and poisoned (\autoref{fig:Poisoned}) configurations. An incorrect moderator temperature could potentially explain this shift, however, the discrepancy persists even when the unpoisoned TMR was warmed to room temperature and vented of methane (\autoref{fig:Unpoisoned_Empty}), where the peak shifts to \qty{85}{meV} experimentally compared to \qty{45}{meV} in simulation\footnote{Measurements were also taken with the unpoisoned TMR at \qty{92}{K}, \qty{100}{K}, and \qty{107}{K} (temperatures accessible within the experimental setup while maintaining liquid methane) but the change in spectra was too small in both experiment and simulation to provide any significant insight.}. This indicates that the spectral shift cannot be attributed solely to methane moderator conditions, including incorrect fill level, methane loss during operation, or errors in the methane thermal scattering models used in simulations.

Systematic errors in either the detector calibration or simulation setup could also contribute to the discrepancy. Potential detector-related issues include incorrect time-channel calibration (the reported \qty{10}{\micro s} channel width may not match the actual electronics setting) or misalignment relative to the neutron extraction channel (alignment was performed with lasers but not quantified), while simulation-related errors could include incorrect material property specifications. These potential systematic effects highlight the importance of calibrating the detector against a known neutron source and benchmarking simulations against reference measurements in future experimental campaigns.

Measurements were also taken at \qty{4.1}{m} and \qty{6.6}{m} and the flux decay with distance was comparable between simulation and experiment, with spectral shapes were consistent across all three measurement positions. This supports that slow neutrons were being measured with distance linearly proportional to flight time and with initial pulse width negligeble to flight time, and that neutron transport beyond the TMR was well-modelled in simulation.

\section{Conclusions} \label{sec:Conclusions}

This paper has presented the design, installation, and initial characterisation of a novel prototype TMR developed as early-stage validation of the design framework for the Versatile ULtra-Compact Accelerator-based Neutron source (VULCAN) project,  which aims to realise a CANS optimised for neutron diffractometry in industrial and university settings. The prototype TMR, with a decoupled liquid methane moderator and configurable in both poisoned and unpoisoned configurations, was tested at CERN's CLEAR facility using a \qty{40}{MeV}, \qty{1.6}{W} electron beam to benchmark simulation predictions of thermal neutron production, transport, and temporal characteristics against experimental measurements.

Neutron spectra recorded using a $^3$He detector demonstrated the detection of thermal neutrons, with approximately \qty{95}{\%} of recorded counts originating from the neutron extraction channel of the TMR. However, significant discrepancies were observed in the measured energy spectra, with a \qty{65}{meV} peak energy experimentally measured for the poisoned and unpoisoned TMRs filled with liquid methane compared to an expected \qty{15}{meV} peak from simulation. This difference in peak energy persisted even when the TMR was warmed to room temperature and vented of methane (\qty{85}{meV} vs \qty{45}{meV}), indicating the cause extends beyond issues with the moderator operation or thermal scattering model accuracy. Other potential contributors include systematic errors in detector time-channel calibration, detector positioning uncertainties relative to the extraction channel, or simulation setup. 

Despite these discrepancies, the measurements establish a foundation for future work. The prototype TMR was successfully fabricated, commissioned, and operated in both configurations, validating the practical aspects of construction, installation, and operation that are also essential to project VULCAN. The installation procedure at CLEAR was derisked, and the facility was demonstrated as suitable for pulsed neutron characterisation experiments.

Recommended next steps include calibrating the $^3$He detector in-situ against a known neutron source to eliminate systematic uncertainties in detector setup. Simulation models can also be benchmarked using the same reference source to check data processing techniques. Additionally, methods to directly monitor the temperature and fill volume of the moderator during operation should be sought, and the effect of detector misalignment with respect to the neutron extraction channel quantified. Future experimental campaigns should incorporate a crystal diffractometer or a chopper-based system to accurately measure the initial FWHM of the generated neutron pulse at specific wavelengths and compare the results between the poisoned and unpoisoned configurations. Additional TMR prototypes incorporating active cooling would enable operation at beam powers closer to the final VULCAN specification ($>$\qty{1}{kW}), while exploration of alternative moderator materials may offer improved performance. Resources are currently being sought to fund these next steps and further advance the VULCAN project.

\begin{acknowledgements}
The authors wish to thank Bruno Guerard, Julien Marchal, Paolo Mutti, and Franck Rey of Institut Laue-Langevin for lending the neutron detectors and assisting with the operation, installation, and data interpretation. The authors are also grateful to the CLEAR operations team for their support and assistance during the measurement campaign.
\end{acknowledgements}

\begin{funding}
We acknowledge funding by the CERN Knowledge Transfer Fund, the CERN Innovation Programme on Environmental Applications (CIPEA), and the EUREKA-EUROSTARS grant (EUROSTARS E! 115722 - VULCAN).
\end{funding}

\ConflictsOfInterest{None.
}

\DataAvailability{The experimental data and code is available from the authors upon reasonable request.}

\bibliographystyle{iucr}   
\bibliography{iucr}  

\end{document}